\documentclass{emulateapj} 
\usepackage{psfig}
\usepackage{mathrsfs}
\usepackage{rotating}
\begin{document}
\slugcomment{ApJ Letters, in press}
\newcommand{\LeeZinn}{\mathscr{L}}

\shortauthors{R. Contreras et al.} 
\shorttitle{M62, An Extremely RR Lyrae-Rich Globular Cluster}

\title{Discovery of $> 200$ RR Lyrae Variables in M62: \\
  An Oosterhoff I Globular Cluster with a Predominantly Blue HB}

\author{R. Contreras, M.~Catelan} 
\affil{Pontificia Universidad Cat\'olica de Chile, Departamento de 
       Astronom\'\i a y Astrof\'\i sica, \\ Av. Vicu\~{n}a Mackenna 4860, 
      782-0436 Macul, Santiago, Chile}
\email{mcatelan@astro.puc.cl} 

\author{Horace A. Smith}  
\affil{Dept.\ of Physics and Astronomy, Michigan State University, 
       East Lansing, MI 48824} 
\email{smith@pa.msu.edu}

\author{Barton J. Pritzl}  
\affil{Macalester College, 1600 Grand Avenue, Saint Paul, MN 55105} 
\email{pritzl@macalester.edu}

\and

\author{J. Borissova}  
\affil{European Southern Observatory, Av. Alonso de C\'ordova 3107, 
  763-0581 Vitacura, Santiago, Chile} 
\email{jborisso@eso.org}

\begin{abstract}
We report on the discovery of a large number of RR Lyrae variable stars 
in the moderately metal-rich Galactic globular cluster M62 (NGC~6266), 
which places it among the top three most RR Lyrae-rich globular clusters 
known. Likely members of the cluster in our studied field, from our 
preliminary number counts, include $\approx 130$ fundamental-mode (RRab) 
pulsators, with $\langle P_{\rm ab}\rangle = 0.548$~d, and $\approx 75$  
first-overtone (RRc) pulsators, with $\langle P_{\rm c}\rangle = 0.300$~d. 
The average periods and the position of the RRab variables with 
well-defined light curves in the Bailey diagram both suggest that the 
cluster is of Oosterhoff type I. However, the morphology of the 
cluster's horizontal branch (HB) is strikingly similar to that of the 
Oosterhoff type II globular cluster M15 (NGC~7078), with a dominant blue 
HB component and a very extended blue tail. Since M15 and M62 differ in 
metallicity by about one dex, we conclude that metallicity, at a fixed 
HB type, is a key parameter determining the Oosterhoff status of a 
globular cluster and the position of its variables in the Bailey diagram.  
\end{abstract}


\section{Introduction}
While Hoyle \& Schwarzschild (1955) were the first to suggest that 
the horizontal branch (HB) represented a post-giant phase in the 
evolution of low-mass stars, it was not until a decade later that 
the first realistic models were computed that provided a convincing 
description of the HB phase in terms of core helium-burning stars 
(Faulkner 1966; Iben \& Faulkner 1968; Castellani \& Renzini 1968; 
Rood \& Iben 1968). 


Since that time, influential papers by Lee, Demarque, \& Zinn (1990), 
Lee (1990, 1991), and Clement \& Shelton (1999) have 
established what is widely perceived to be the current paradigm in 
the area, which consists of the following main ingredients: 
i)~RR Lyrae (RRL) variables in Oosterhoff (1939) type II (OoII) GCs, 
with their predominantly blue HB types, are evolved away from a 
position on the blue zero-age HB; 
ii)~Conversely, RRL stars in OoI clusters, with their uniformly 
populated or red HB morphologies, are mostly ``unevolved'' objects; 
iii)~The position of stars on the Bailey (period-amplitude) diagram, 
and therefore the Oosterhoff type, is determined not by the metallicity 
of the cluster, but rather by the HB type. In most of the quoted works, 
``HB type'' is usually taken to be the parameter $\LeeZinn = (B-R)/(B+V+R)$, 
where $B$, $R$, and $V$ are the numbers of blue, red, and variable 
(RRL) HB stars, respectively---though many alternative definitions 
also exist (e.g., Fusi Pecci et al. 1993). 

On the other hand, several authors have cautioned that some key 
observational features involving both variable and non-variable HB 
stars may not be straightforwardly accounted for in this way (e.g., 
Rood \& Crocker 1989; Fusi Pecci \& Bellazzini 1997; Pritzl et al. 
2002; Sweigart \& Catelan 1998; Catelan 2004a,b). Accordingly, 
additional observational tests of state-of-the-art theoretical 
HB models that might help unveil missing or inadequate ingredients 
in the calculations may prove of relevance for further progress 
in the area. 

In this sense, little attention has been paid so far to the case of 
Galactic GCs with predominantly {\em blue} HB types, significant 
quantities of RRL variables, but apparently of {\em OoI} type. 
Indeed, according to the compilation of Castellani \& Quarta (1987), 
several such cases do exist, perhaps most notably those of NGC~4147, 
M14 (NGC~6402), and M62 (NGC~6266). 

On the other hand, variability surveys for most such GCs have  
generally been carried out several decades ago, and are necessarily 
incomplete. In addition, it has been noted (Clement 2000) that 
the reported RRL periods in some of these studies are incorrect, 
which may render their classification into an Oosterhoff type suspect. 

In order to clarify the situation, we have started a program to search 
for previously unknown variable stars in the aforementioned clusters,
on the basis of state-of-the-art detection and analysis techniques. 
We chose to start this project with M62, which is an extremely 
massive GC ($M_V = -9.19$; Harris 1996) with the 
largest known number of RRL variables (74, according to the 
latest revision to the Clement et al. 2001 catalog) among these 
clusters, thus giving us the best possibility to 
study a statistically significant sample of variable stars in a 
region of parameter space not previously covered in the literature. 

Previous surveys for variable stars in M62 were carried out by van 
Agt \& Oosterhoff (1959) and Gascoigne \& Ford (1967), revealing (as 
already mentioned) a rich harvest of RRL variables. Studies of 
the color-magnitude diagram (CMD) of the cluster, on the other hand, 
have revealed the presence of an HB morphology which is strikingly 
similar to that of M15 (NGC~7078) (Caloi, Castellani, \& Piccolo 1987; 
Piotto et al. 2002), and certainly much bluer than M68's (NGC~4590). 
On the basis of the Piotto et al. HST diagrams, 
without applying completeness corrections at the faint end of the 
blue HB distribution, we obtain $\LeeZinn \simeq 0.55$, and  
$\LeeZinn \simeq 0.65$ for M15. Other HB morphology parameters that 
are more sensitive to the detailed properties of the blue HB (e.g., 
Fusi Pecci et al. 1993) are likely to be even more similar between 
the two clusters. M15 ($M_V = -9.17$; Harris 1996) 
has essentially the same mass as M62, and both clusters are 
suspected of having collapsed cores. The chief difference between 
the two is their metallicities, with the Harris catalog 
indicating a value ${\rm [Fe/H]} = -1.29$ for M62 (comparable to 
that of M5~=~NGC~5904) but a much lower ${\rm [Fe/H]} = -2.26$ for 
M15. Which of these effects dominate the determination of the 
Oosterhoff type of the cluster---its moderately high metallicity, 
which would favor an OoI type (e.g., Sandage, Katem, \& Sandage 
1981); or its M15-like blue HB 
morphology, which would suggest instead an OoII classification? 

The main result of our survey, which we report on in this {\em Letter}, 
is that metallicity is the dominant factor in the case of M62, since we 
confirm, with the discovery of over 200 variables in this cluster, the 
original indications (Castellani \& Quarta 1987) that M62 is indeed 
OoI, in spite of its predominantly blue HB.

\section{Observations and Data Reduction}
Time-series observations in $B$ and $V$ were obtained by two of us 
(RC and JB) with the Warsaw 1.3m telescope at the Las Campanas
Observatory (LCO), in the course of 7 consecutive nights over the period 
April 6--13 2003. The camera used is the 8kMOSAIC camera, comprised 
of eight $2040\times 4096$ chips, with a scale $0.26\arcsec/{\rm pixel}$. 
A total of 126 images in $B$ and 126 in $V$ were secured with this 
setup. 

An additional 42 images in $B$ and 42 images in $V$ were also obtained 
with the Cerro Tololo Inter-American Observatory (CTIO) 1.3m telescope 
in service mode, using the Andicam $1024\times 1024$ CCD, with a scale 
$0.369\arcsec/{\rm pixel}$, over the timespan April 24 2003 to June 30 2003. 

The LCO images were pre-processed with the Warsaw 1.3m pipeline, so 
that no additional pre-reduction steps were necessary. The CTIO images 
were pre-reduced using standard procedures. 
 
To perform the photometry and search for variable stars, we have employed 
both the image-subtraction package ISIS v2.1 (Alard \& Lupton 1998; 
Alard 2000) and ALLFRAME (Stetson 1994). So far we have only analyzed 
the Warsaw 1.3m chip where the bulk of the cluster stars are located, 
in addition to the CTIO images. A full description of our adopted 
procedures, as well as the calibration, will be described in a 
forthcoming paper (Contreras et al. 2005). While the calibration that 
is used in the present work is not yet final, we believe it should be 
good enough to at least yield pulsation amplitudes that are reliable 
to within a few hundredths of a magnitude.

\section{Results}
While we chose M62 for the present study because of its previously known 
``large'' RRL population (74 variables), and while we were well 
aware that a significant increase in the number of known variables 
was likely in store due to the usage of state-of-the-art detection 
and analysis techniques (see, e.g., Corwin et al. 2004 and references 
therein), our survey revealed, much to our surprise, that M62 is in 
fact one of the most RRL-rich of all known GCs. The total number 
of RRL variables detected so far in our survey is 212, the 
vast majority of which, according to our preliminary analysis, is 
likely to be cluster members. We note that there are 
an additional 17 RRL variables listed in the Clement et al. (2001) 
catalog which fall outside the fields that we have analyzed so far, 
thus indicating that the total RRL population of the cluster 
could be even larger. For comparison, M3 (NGC~5272) is currently the 
most RRL-rich GC known, with a total of about 230 
identified variables (Clementini et al. 2004); followed (according to 
the Clement et al. catalog) by $\omega$~Centauri (NGC~5139), with a 
total of 161 RRL; M5 (NGC~5904), with $N_{\rm RR} = 126$; and M15, 
with $N_{\rm RR} = 88$. Our detections clearly place M62 among the top 
three clusters in terms of known RRL variables, and it is not 
impossible that more extensive analyses of the innermost regions of 
the cluster and neighboring fields will reveal it to be the most 
RRL-rich of all known Galactic globulars. 
We compute a specific frequency of RRL 
variables $S_{\rm RR} = N_{\rm RR}\,10^{0.4 (7.5+M_V)} \approx 46$, 
which is higher than all but nine 
clusters according to the Harris (1996) catalog. In Figure~1, we 
show a few sample light curves for fundamental-mode (RRab) variables 
detected in our survey; whereas Figure~2 shows sample RRc light 
curves.

\begin{figure}[t]
  \figurenum{1}
  \epsscale{1}
\plotone{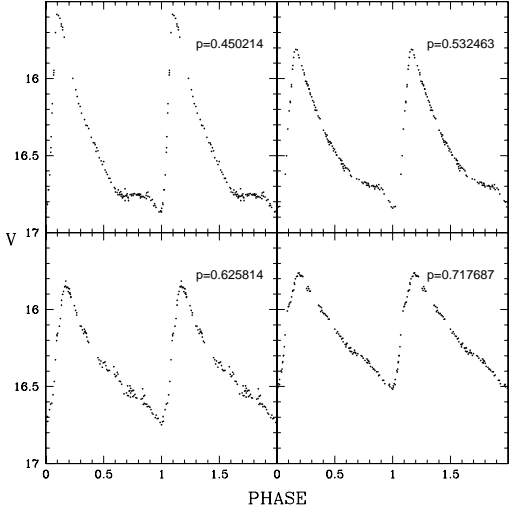}
  \caption{Light curves for some of the RRab variables identified in 
    our survey.   
      }
      \label{Fig01}
\end{figure}

\begin{figure}[ht]
  \figurenum{2}
  \epsscale{1}
\plotone{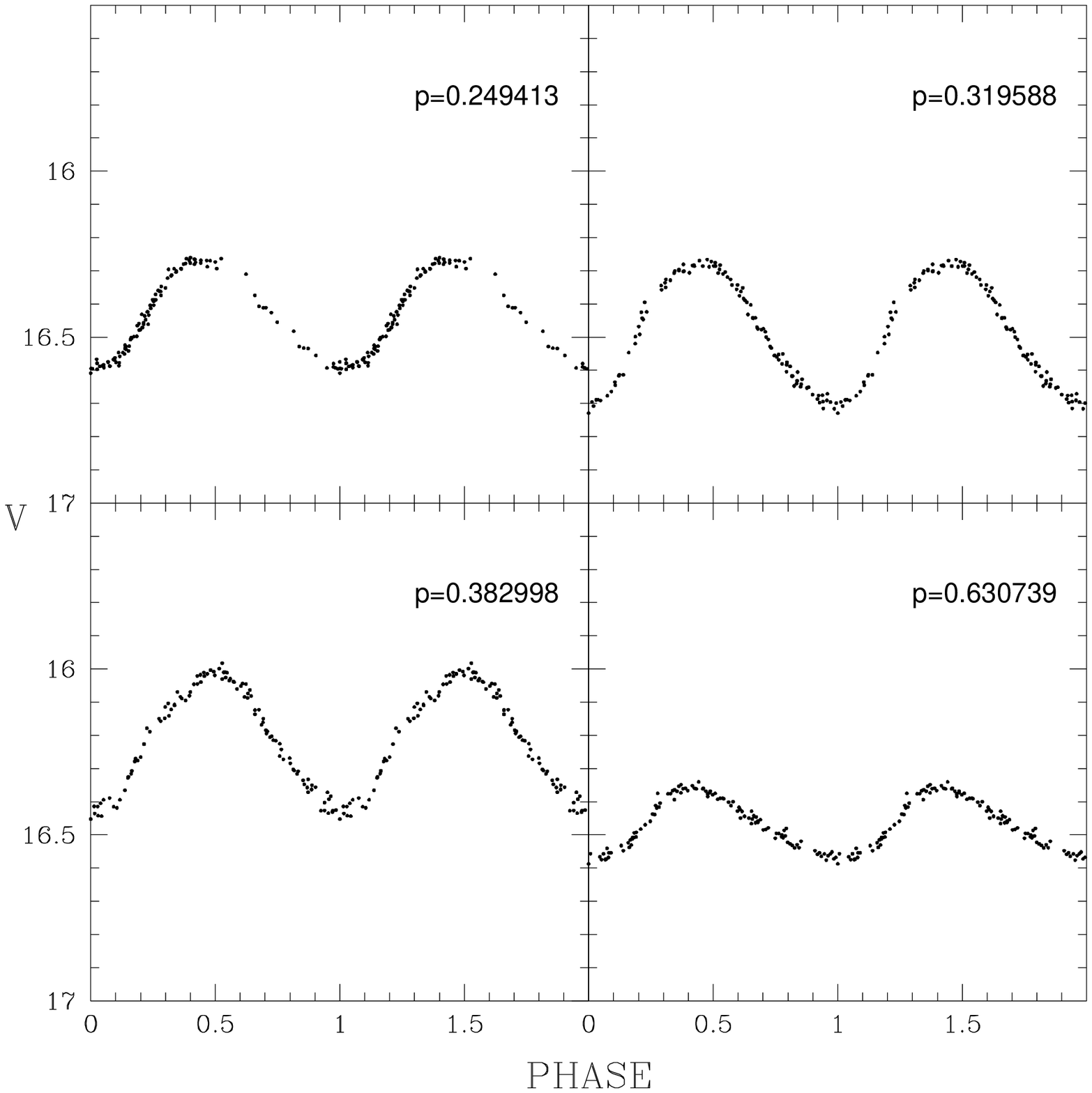}
  \caption{Light curves for some of the RRc identified in our survey. 
    Note the possible presence of a first-overtone pulsator with an 
    extremely long period (lower right plot).  
      }
      \label{Fig02}
\end{figure}

Note the possible presence of an RRc with $P > 0.6$~d in Figure~2. 
That the star is indeed an RRc is suggested by several plots that 
make use of Fourier decomposition of the light curve, particularly 
those involving the amplitude ratio $A_{21}$. However, it does not 
appear to occupy the usual locus of long-period RRc stars in the 
period-$\phi_{31}$ plane. Catelan (2004b) notes that 
long-period RRc's may be peculiarly bright, but there is no clear 
indication, from the CMD, that this is the case with this star 
(though differential reddening could in principle account for this 
discrepancy). We have found no evidence that the star may be 
a blend. There is a second example in M62, for which however 
we have not yet been able to obtain a calibrated light curve. 
Long-period RRc stars (i.e., RRc with $P > 0.45$~d) are not only 
exceedingly rare, with only a handful of examples currently 
known in $\omega$~Cen, NGC~6388, NGC~6441, and M3 (Catelan 2004b and 
references therein); they had also never previously been identified 
with periods longer than about 0.57~d. It remains to be seen whether 
these stars will extend the currently known period domain for 
first-overtone RRL.

\begin{figure}[ht]
  \figurenum{3}
  \epsscale{1}
\plotone{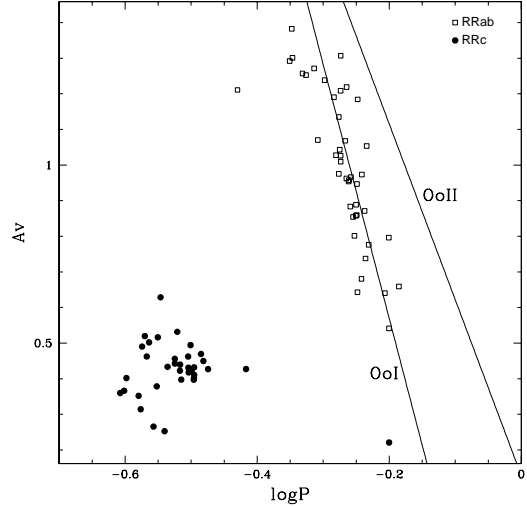}
  \caption{Bailey diagram for M62 RRL stars. RRab are shown as 
    open squares, and RRc as filled circles. Only RRab with $D_m < 5$ 
    are shown. 
      }
      \label{Fig03}
\end{figure}

\subsection{Oosterhoff Classification} 
After removing RRL variables which are clearly non-members, 
and also variables with uncertain periods, we obtain $\approx 130$ 
RRab and $\approx 75$ RRc with reliable period determinations. 
Based on these stars, we find $\langle P_{\rm ab}\rangle = 0.548$~d, 
and $\langle P_{\rm c}\rangle = 0.300$~d. These values are clearly 
consistent with the classification of the cluster into type OoI 
(e.g., Smith 1995), thus fully confirming the previous classification 
(Castellani \& Quarta 1987), which however was based on many fewer 
stars and more uncertain light curves. The variable stars in the 
Clement et al. (2001) catalog which fall outside our studied fields 
were not included when computing these numbers. 

The Oosterhoff classification of the cluster can be checked by 
verifying the position of the RRab variables in the Bailey 
(period-amplitude) diagram. As is well known, the position of a star 
in this diagram can be strongly affected by the presence of the 
Blazhko effect. In order to minimize this problem, we have computed 
the Jurcsik \& Kov\'acs (1996) $D_m$ parameter, based on Fourier 
decomposition, for all variables with calibrated light curves. 
According to several authors (Clement \& Shelton 1999 and references 
therein; but see also Cacciari, Corwin, \& Carney 2005), small $D_m$ 
values provide a good criterion for cleaning the sample from Blazhko 
variables and other variables with problematic light curves. 
Accordingly, we limit our analysis to stars with $D_m < 5$. 
 
In Figure~3 we show the position of the RRab with $D_m < 5$ in the 
$A_V - \log P$ diagram. RRc variables are also indicated, and occupy 
the lower left region of the diagram as usual. Overplotted are the 
Clement (2000, priv. comm.) lines for OoI and OoII stars. In spite 
of the presence of some residual scatter, particularly at the 
short-period end of the RRab distribution, it is clear that the 
Bailey diagram confirms the classification of the cluster into 
an OoI type. 

The one intriguing aspect of M62 which is not clearly consistent with 
its OoI classification is the large number fraction of RRc variables. 
From the above number counts, we find 
$f_c = N_{\rm c}/N_{\rm RR} \simeq 0.36$. Note that this 
represents a large upward revision with respect to the previously 
available number counts, which gave instead $f_c \simeq 0.16$ 
(Castellani \& Quarta 1987)---clearly, our new variability survey 
has preferentially discovered new RRc stars, which is not surprising 
given the generally smaller amplitude of their light curves. While 
we note that the M62 c-to-ab number ratio is more in line with the 
cluster's blue HB morphology than with its OoI classification, it 
should also be noted that there exist other OoI clusters with 
relatively large RRc number fractions, such as NGC~6171 (M107), 
whose $f_c$ is essentially identical to M62's; and NGC~4147, which 
has more RRc than RRab variables (see also Arellano Ferro et al. 2004). 

\begin{figure}[t]
  \figurenum{4}
  \epsscale{1}
\plotone{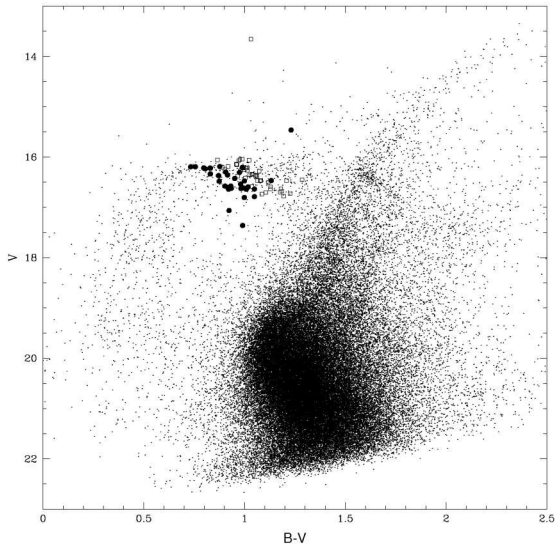}
  \caption{M62 CMD, based on our ALLFRAME reductions of 100 selected 
    images. RRc variables are indicated by filled circles, and RRab 
    by open squares.  
      }
      \label{Fig04}
\end{figure}

\subsection{Color-Magnitude Diagram} 
As a result of our ALLFRAME reductions, we were able to obtain a deep 
CMD for M62, which however is severely affected by differential 
reddening in this low Galactic field ($l = 353.57\degr$, 
$b = 7.32\degr$), consistent with the cluster's fairly large 
reddening of $E(\bv) = 0.47$ (Harris 1996). We show a calibrated CMD 
of the cluster in Figure~4, where stars with the ALLFRAME photometry 
parameter $\chi \geq 1.5$ and with 
large frame-to-frame dispersion (except for confirmed RRL variables) 
were removed. We have checked that the ``second red giant branch" to the 
right of the main branch is an artifact of patchy extinction accross the 
face of the cluster, and is indeed absent in the HST CMD of Piotto et 
al. (2002). RRL variables are also shown in the plot, 
with filled circles (RRc) and open squares (RRab). Only variables with 
sufficiently well-defined light curves that Fourier decomposition could 
be reliably performed and average magnitudes computed accordingly were 
included. Note how nicely the instability strip blue and red edges 
slope along the direction of the reddening vector. The amplitude ratios 
indicate that the red and overluminous RRc is indeed a blend with a 
red star. A detailed investigation of the CMD position of all our 
detected variables, which also includes a large number of type II 
Cepheids, will be left for a future paper (Contreras et al. 2005).

\section{Summary and Discussion} 
In this {\em Letter}, we were able to show that M62 is one of the most 
RRL-rich GCs known, with a total number of RRL stars that is comparable 
to (and may eventually exceed) that in M3. Moreover, we have shown that, 
in spite of a blue HB that bears striking resemblance to M15's and is 
much bluer than M68's (perhaps the two most famous OoII clusters), 
the cluster is instead of type OoI, as 
confirmed both by the average periods of its RRab and RRc variables 
and by the position of the RRab stars in the Bailey diagram. Given 
that the main difference between M62 and M15 is the fact that the latter
is more metal-poor than the former by about 1~dex, we conclude that 
metallicity (in addition to HB type) is an important factor in defining 
the position of RRab 
stars in the Bailey diagram, as well as in defining Oosterhoff status. 
Whereas, based on the current paradigm in the evolutionary interpretation 
of HB stars in GCs, one might have expected an OoII classification for 
this cluster (e.g., Clement 2000), it should be noted that HB theory does 
predict that it should be progressively more difficult to produce 
significant numbers of RRL stars evolved from a position on the blue 
zero-age HB with increasing metallicity (Pritzl et al. 2002). Metallicity,   
in addition to HB type, should accordingly play a non-neglibible role---and 
this is indeed confirmed by our results.

\acknowledgments We thank C. Clement, R. Salinas, A. W. Stephens, 
P. B. Stetson, and M. Zoccali for useful discussions and/or information. 
M.C. and R.C. acknowledge support by Proyecto FONDECYT Regular No. 1030954. 
B.J.P. would like to thank the National Science Foundation (NSF) for support 
through a CAREER award, AST 99-84073. H.A.S. acknowledges the NSF for support 
under grant AST 02-05813.

\end{document}